\documentclass[twocolumn,secnumarabic,amssymb,letter,nobibnotes, aps, prd]{revtex4-1}
\usepackage{graphicx}
\usepackage{amsmath}
\usepackage{amsfonts}
\usepackage{amssymb}

\newcommand{\dif}{\mathrm{d}}

\begin{document}
\title{\Large\bf {Thermodynamics of the in-medium chiral condensate}\footnote{
Work supported in part by BMBF, GSI and by the DFG Excellence Cluster ``Origin and 
Structure of the Universe''.}}
\vskip 1.5true cm
\author{S. Fiorilla, N. Kaiser  and W. Weise}
\affiliation{Physik-Department, Technische Universit\"{a}t M\"{u}nchen, D-85747
Garching, Germany}
\begin{abstract}
The temperature dependence of the chiral condensate in isospin-symmetric nuclear 
matter at varying baryon densities is investigated starting from a realistic free 
energy density of the correlated nuclear many-body system. The framework is thermal
in-medium chiral effective field theory which permits a systematic calculation of 
the quark mass dependence of the free energy density. One- and two-pion exchange 
processes, virtual $\Delta(1232)$-isobar excitations and  Pauli blocking 
corrections are treated up to and including three-loop order. It is found that 
nuclear matter remains in the Nambu-Goldstone phase with spontaneously broken 
chiral symmetry at least in the range of temperatures $T\lesssim 100\,$MeV and 
baryon densities up to about twice the density of normal nuclear matter. 
\end{abstract}
\maketitle


The chiral condensate $\langle \bar{q}q \rangle$, i.e. the expectation value of 
the scalar quark density, plays a fundamental role as an order parameter of 
spontaneously broken chiral symmetry in the hadronic low-energy phase of QCD. The 
variation of $\langle \bar{q}q \rangle$ with temperature and baryon density is a 
key issue for locating the chiral transition boundary in the QCD phase diagram. 
The melting of the condensate at high temperatures and/or densities determines the 
crossover from the Nambu-Goldstone phase to the Wigner-Weyl realization of chiral 
symmetry in QCD. 

It is thus of principal interest to perform a systematically organized
calculation of the thermodynamics of the chiral condensate. Such a calculation 
requires knowledge of the free energy density as a function of the quark mass (or 
equivalently, as a function of the pion mass). The appropriate framework for such 
a task is in-medium chiral effective field theory with its explicit access to one- 
and two-pion exchange dynamics and the resulting two- and three-body correlations 
in the presence of a dense nuclear medium. 

Previous studies of the in-medium variation of the chiral condensate were
mostly concerned with the density dependence of  $\langle \bar{q}q \rangle$ at 
zero temperature, using different approaches such as QCD sum rules \cite{drukarev} 
or models \cite{LiKo94,bw96} based on the boson exchange phenomenology of nuclear 
forces. Temperature effects have been included in schematic Nambu - Jona-Lasinio 
(NJL) approaches \cite{HK,KLW}. Such NJL models work with quarks as quasiparticles 
and provide useful insights into dynamical mechanisms behind spontaneous chiral 
symmetry breaking and restoration, but  they do not properly account for nucleons 
and their many-body correlations, a prerequisite for a more realistic treatment.    

The present work extends a recent chiral effective field theory calculation 
\cite{cond} of the density-dependent in-medium  $\langle \bar{q}q \rangle$ 
condensate to finite temperatures $T$. A related chiral effective field theory 
approach to nuclear matter at $T=0$ performing resummations to all orders has 
been reported in ref.\cite{oller}. Corrections to the linear density 
approximation are obtained by differentiating the interaction parts of the free 
energy density of isospin-symmetric nuclear matter with respect to the (squared) 
pion mass. Effects from one-pion exchange (with $m_\pi$-dependent 
vertex corrections), iterated $1\pi$-exchange, and irreducible $2\pi$-exchange 
including intermediate $\Delta(1232)$-isobar excitations, with Pauli-blocking 
corrections up to three-loop order are systematically treated. The dominant nuclear 
matter effects on the dropping condensate are supplemented by a further small 
reduction due to interacting thermal pions. To anticipate the result: we find that 
the delayed tendency towards chiral symmetry restoration with increasing baryon 
density $\rho$, observed at $T=0$ \cite{cond} in the same framework, gets gradually 
softened with increasing temperature. An approximately linear decrease of the 
quark condensate with increasing $\rho$ is recovered at temperatures around 
$T\simeq 100\,$MeV. However, no rapid drive towards a first order chiral phase 
transition is seen, at least up to $\rho \lesssim 2\,\rho_0$ where
$\rho_0 = 0.16$ fm$^{-3}$ is the density of normal nuclear matter.   

Consider the free energy density ${\cal F}=\rho \bar F(\rho,T)$ of 
isospin-symmetric (spin-saturated) nuclear matter, with $\bar{F}(\rho,T)$ the free 
energy per particle. In the approach to nuclear matter based on in-medium chiral 
perturbation theory \cite{nucmatt,deltamat}  the free energy density is given by a 
sum of convolution integrals of the form, 
\begin{eqnarray} \rho \, \bar
F(\rho,T)&=& 4\int_0^\infty \!\! \dif p\, p \, {\cal K}_1\,d(p)\nonumber\\ 
&&+\int_0^\infty \!\!\dif p_1\int_0^\infty \!\!\dif p_2\, {\cal K}_2\, d(p_1)
d(p_2)\nonumber \\ &&+ \int_0^\infty \!\!\dif p_1\int_0^\infty \!\!\dif
p_2\int_0^\infty \!\! \dif p_3\, {\cal K}_3 \,d(p_1) d(p_2)d(p_3)\nonumber\\ 
&&+\rho \, \bar {\cal A}(\rho,T)\,, \end{eqnarray}
where ${\cal K}_1, {\cal K}_2$ and ${\cal K}_3$ are one-body, two-body and 
three-body kernels, respectively. The last term, the so-called anomalous 
contribution $\bar {\cal A}(\rho,T)$ is a special feature at finite 
temperatures \cite{kohn} with no counterpart in the calculation of 
the groundstate energy density at $T=0$. As shown in ref.\cite{nucmatt} the 
anomalous contribution arising in the present context from second-order 
pion exchange has actually very little influence on the equation of state of 
nuclear matter at moderate temperatures $T< 50\,$MeV. 

The quantity  
\begin{equation} 
d(p) = {p\over 2\pi^2} \bigg[ 1+\exp{p^2/2M_N -\tilde \mu \over T} 
\bigg]^{-1} \,,
\end{equation}
denotes the density of nucleon states in momentum space. It is the product of 
the temperature dependent Fermi-Dirac distribution and a kinematical prefactor 
$p/ 2\pi^2$ which has been included in $d(p)$  for convenience. 
$M_N$ stands for the (free) nucleon mass. The particle density $\rho$ is 
calculated as
\begin{equation}
\rho= 4\int_0^\infty  \!\! \dif p\, p \,d(p) \,. \end{equation} 
This relation determines the dependence of the effective one-body 
chemical potential $\tilde \mu(\rho,T;M_N)$ on the thermodynamical variables 
$(\rho, T)$ and indirectly also on the nucleon mass $M_N$. 
The one-body kernel ${\cal K}_1$ in eq.(1) provides the contribution of the
non-interacting nucleon gas to the free energy density and it reads 
\cite{nucmatt}
\begin{equation} \label{K1}
{\cal K}_1 = M_N +\tilde \mu- {p^2\over 3M_N}- {p^4\over 8M_N^3} \,. 
\end{equation}
The first term in ${\cal K}_1$ gives the leading contribution (density $\rho$
times nucleon rest mass $M_N$) to the free energy density. The 
remaining terms account for (relativistically improved) kinetic energy corrections.
 
Our starting point is the Feynman-Hellmann theorem which relates the 
temperature dependent in-medium quark condensate $\langle \bar q q\rangle(\rho,
T)$ to the quark mass derivative of the free energy density of 
isospin-symmetric (spin-saturated) nuclear matter. Using the 
Gell-Mann-Oakes-Renner relation $m_\pi^2 f_\pi^2 = -m_q \langle 0|\bar q q|0
\rangle$ one finds for the ratio of the in-medium to vacuum quark condensate
\begin{equation}  
{\langle \bar q q\rangle(\rho,T)\over  \langle 0|\bar q q|0
\rangle} = 1 - {\rho \over f_\pi^2} {\partial \bar F(\rho,T) \over \partial 
m_\pi^2} \,,\end{equation}
where the derivative with respect to $m_\pi^2$ is to be taken at fixed $\rho$ 
and $T$. The quantities  $\langle 0|\bar q q|0\rangle $ (vacuum quark 
condensate) and $f_\pi$ (pion decay constant) are to be understood as 
taken in the chiral limit, $m_q\to 0$. Likewise, $m_\pi^2$ stands for the leading 
linear term in the quark mass expansion of the squared pion mass. 

In the one-body kernel  ${\cal K}_1$ the quark (or pion) mass 
dependence is implicit via its dependence on the nucleon mass $M_N$. The 
condition $\partial \rho/\partial M_N=0$ applied to eq.(3) leads to the 
following dependence of the effective one-body chemical potential $\tilde \mu$ 
on the nucleon mass $M_N$:
\begin{equation} {\partial \tilde \mu \over \partial M_N }= {3 \rho \over 2M_N 
\Omega_0''} \,, \qquad   \Omega_0''= -4M_N  \int_0^\infty  \!\! \dif p\, 
{d(p) \over p}\,. \end{equation} 
The nucleon sigma term $\sigma_N = \langle N|m_q \bar q q |N\rangle= m_\pi^2 \,
\partial M_N/\partial m_\pi^2$ measures the variation of the nucleon mass $M_N$ with
the quark (or pion) mass. Combining both relationships leads to the 
$m_\pi^2$-derivative of the one-body kernel:
\begin{equation} {\partial {\cal K}_1\over \partial m_\pi^2} = {\sigma_N \over 
m_\pi^2} \bigg\{ 1+ {3 \rho \over 2M_N \Omega_0''} +{p^2 \over 3M_N^2} +{3p^4 
\over 8M_N^4} \bigg\}\,. \end{equation}
In the limit of zero temperature, $T=0$, the terms in eq.(7) reproduce the 
linear decrease of the chiral condensate with density. The kinetic 
energy corrections account for the (small) difference between the scalar and 
the vector (i.e. baryon number) density. In the actual calculation we  
use the chiral expansion of the nucleon sigma term $\sigma_N$ to order 
${\cal O}(m_\pi^4)$ as given in eq.(19) of ref.\cite{cond}. The empirical value 
of the nucleon sigma term (at the physical pion mass $m_\pi = 135\,$MeV) is 
$\sigma_N =(45\pm 8)\,$MeV \cite{gls}. In numerical calculations we choose the 
central value.

The two- and three-body kernels, ${\cal K}_2$ and ${\cal K}_3$, 
related to one-pion exchange and iterated one-pion exchange have already been given 
in explicit form in ref.\cite{nucmatt}. Their derivatives with respect to the 
squared pion mass, $\partial {\cal K}_{2,3}/\partial m_\pi^2$, are hence obvious  
and do not need to be written out here. The same applies to the anomalous 
contribution $\bar {\cal A}(\rho,T)$ (see eqs.(14,15) in ref.\cite{nucmatt}) and
to the two- and three-body kernels related to $2\pi$-exchange with excitation 
of virtual $\Delta(1232)$-isobars (see section 6 in ref.\cite{deltamat}). In 
case of the one-pion exchange contribution we include the $m_\pi$-dependent 
vertex correction factor $\Gamma(m_\pi)$ as discussed in section 2.1 of 
ref.\cite{cond}. The short-distance contact term which produces a 
$T$-independent correction of order $\rho^2$ to the in-medium condensate is 
treated  exactly in the same way as in ref.\cite{cond}, i.e. all terms with a 
non-analytical quark-mass dependence generated by pion-loops are taken into 
account.   
\begin{figure}
\includegraphics*[totalheight=2.cm]{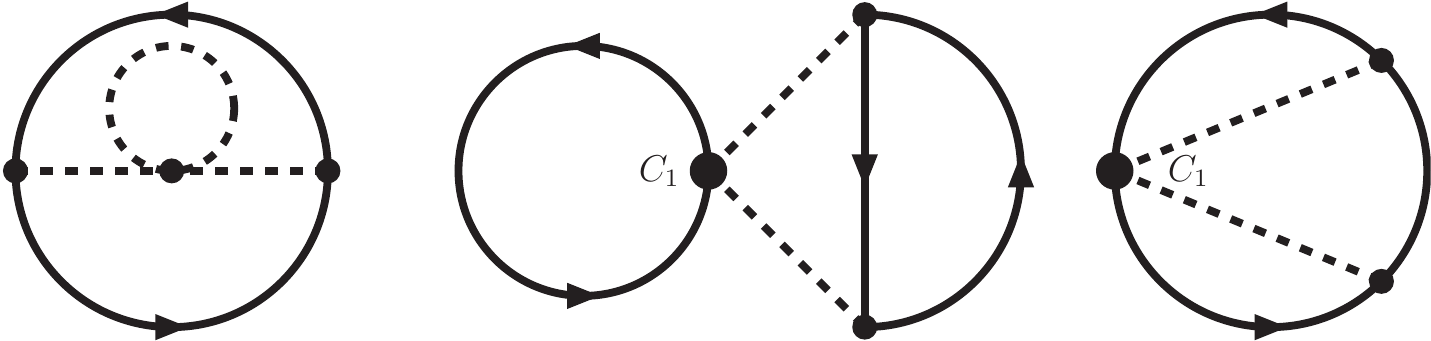}
\caption{Three-loop contributions to the free energy density of nuclear matter 
that are relevant for the in-medium chiral condensate. Left diagram:
pion self-energy correction; central and right diagram: two-pion exchange Hartree 
and Fock terms involving the $\pi\pi NN$ contact interaction proportional to the 
low-energy constant $c_1m_\pi^2$.}
\label{figure1}
\end{figure}
 
Let us now turn to some three-loop contributions which are new and of special 
relevance for the in-medium chiral condensate. The first one comes from the pion 
selfenergy diagram shown in Fig.\,1. It gives rise to the following derivative of 
the two-body kernel:
\begin{eqnarray} 
{\partial {\cal K}_2^{(\pi)}\over \partial m_\pi^2 }
&=& {3g_A^2 m_\pi^2 \over \pi^2 (4f_\pi)^4} \Big\{ \bar{\ell}_3\big[(X^-_{12})^2 -
(X^+_{12})^2\big] + (4\bar{\ell}_3 - 1)\nonumber\\ && \times \left(X^+_{12} - 
X^-_{12} \right)+ (2\bar{\ell}_3 - 1)\ln{X^-_{12}\over X^+_{12}}\Big\}\,,\end{eqnarray}
with the abbreviations $X^\pm_{ij} = [1 + (p_i\pm p_j)^2/m_\pi^2]^{-1}$ and the 
$\pi\pi$ low-energy constant $\bar{\ell}_3 \simeq 3$. 

The chiral $\pi\pi NN$ contact vertex proportional to $c_1 m_\pi^2 $ generates 
$2\pi$-exchange Hartree and Fock diagrams, also shown in Fig.1. Concerning the 
free energy density $\rho \bar F(\rho,T)$ or the equation of state of nuclear 
matter their contributions are actually almost negligible. However,
when taking the derivative with respect to $m_\pi^2$ as required for the 
calculation of the in-medium condensate, these contributions turn out to be of 
similar importance as other interaction terms. The corresponding contribution to 
the derivative of the two-body kernel reads:
\begin{equation} {\partial {\cal K}_2^{(c_1)}\over \partial m_\pi^2 } = {g_A^2 
c_1 m_\pi^3 \over 8\pi f_\pi^4} \bigg\{ G\Big({p_1+p_2 \over 2m_\pi} \Big)- 
 G\Big({p_1-p_2 \over 2m_\pi} \Big) \bigg\}\,,  \end{equation}
with the auxiliary function:
\begin{equation} G(x)=8x(3+x^2)\arctan x -5\ln(1+x^2)-100 x^2\,.\end{equation}
The $2\pi$-exchange Hartree diagram with one $c_1m_\pi^2$-vertex contributes
the following piece to the three-body kernel: 
\begin{equation}  
{\partial {\cal K}_3 ^{(c_1)}\over \partial m_\pi^2}=  {6g_A^2 c_1 p_3\over f_\pi^4} 
\Big\{(X^+_{12}-X^-_{12})(X^+_{12}+X^-_{12}-3) +\ln {X^+_{12}\over X^-_{12}} \Big\},
\end{equation}
while the three-body term associated with the  $2\pi$-exchange Fock diagram with 
one $c_1m_\pi^2$-vertex gives:
\begin{eqnarray} {\partial {\cal K}_3^{(c_1)}\over \partial m_\pi^2 } 
&=&  {3g_A^2 c_1  \over f_\pi^4} \bigg[{p_2\over p_3} +{p_3^2-p_2^2-m_\pi^2 \over 
4p_3^2} \ln {X^-_{23}\over X^+_{23}}  \bigg] \nonumber \\ && \times \bigg[
\,p_1+{p_3^2-p_1^2-3m_\pi^2 \over 4p_3} \ln {X^-_{13} \over X^+_{13}}\nonumber\\  && 
+(p_1+p_3)X^+_{13} +(p_1-p_3) X^-_{13}\, \bigg]\,. \end{eqnarray}

Last not least we incorporate the effects of thermal pions. Through its  
$m_\pi^2$-derivative the pressure (or free energy density) of thermal pions gives 
rise to a further reduction of the $T$-dependent in-medium condensate. In the 
two-loop approximation of chiral perturbation theory including effects from the 
$\pi\pi$-interaction one finds the following shift of the condensate ratio in the 
presence of the pionic heat bath \cite{gerber,toublan,pipit}: 
\begin{eqnarray}&&{\delta\langle \bar q q\rangle(T)\over  \langle 0|\bar q q|0
\rangle} =-{3m_\pi^2 \over (2\pi f_\pi)^2} H_3\Big({m_\pi \over T}\Big) \bigg\{1+ 
{m_\pi^2 \over 8\pi^2 f_\pi^2} \nonumber \\ & &\qquad \quad \times \bigg[H_3\Big(
{m_\pi \over T}\Big)-  H_1\Big({m_\pi \over T}\Big) + {2-3\,\bar{\ell}_3 \over 8} 
\bigg] \bigg\},  \end{eqnarray}
with the functions $H_{1,3}(m_\pi/T)$ defined by  integrals over the 
Bose distribution of thermal pions:
\begin{eqnarray}
 H_1(y) &=& \int_y^\infty \!\! dx\, {1 \over \sqrt{x^2-y^2}
(e^x-1)}\,,\nonumber\\  H_3(y) &=& y^{-2} \int_y^\infty \!\! dx\, 
{\sqrt{x^2-y^2} \over e^x-1}\,.
\end{eqnarray}

We proceed with a presentation of results. As input we consistently use the same 
parameters in the chiral limit as in our previous works \cite{cond}, namely: 
$f_\pi = 86.5\,$MeV, $g_A =1.224$, $c_1 =-0.93\,$GeV$^{-1}$ and $M_N = 882\,$MeV. 
Concerning the contact term representing unresolved short-distance dynamics, we 
recall from ref.\,\cite{cond} that its quark mass dependence, estimated from recent 
lattice QCD results for the nucleon-nucleon potential \cite{hatsuda}, is 
negligibly small compared to that of the intermediate and long range 
(pion-exchange driven) pieces. 

It is worth pointing out that in-medium chiral perturbation theory with this input 
produces a realistic nuclear matter equation of state \cite{deltamat}, including a 
proper description of the thermodynamics of the (first-order) liquid-gas phase 
transition. Apart from temperature $T$, the additional ``small" parameter in this 
approach is the nucleon Fermi momentum $p_F$ in comparison with the chiral scale, 
$4\pi f_\pi \sim 1$\,GeV. Our three-loop calculation of the free energy density is 
reliable up to about 
twice the density of normal nuclear matter.  It can be trusted over a temperature 
range (up to $T\sim 100$ MeV) in which the hot and dense matter still remains well 
inside the phase of spontaneously broken chiral symmetry. 

\begin{figure}
\includegraphics*[totalheight=6.3cm]{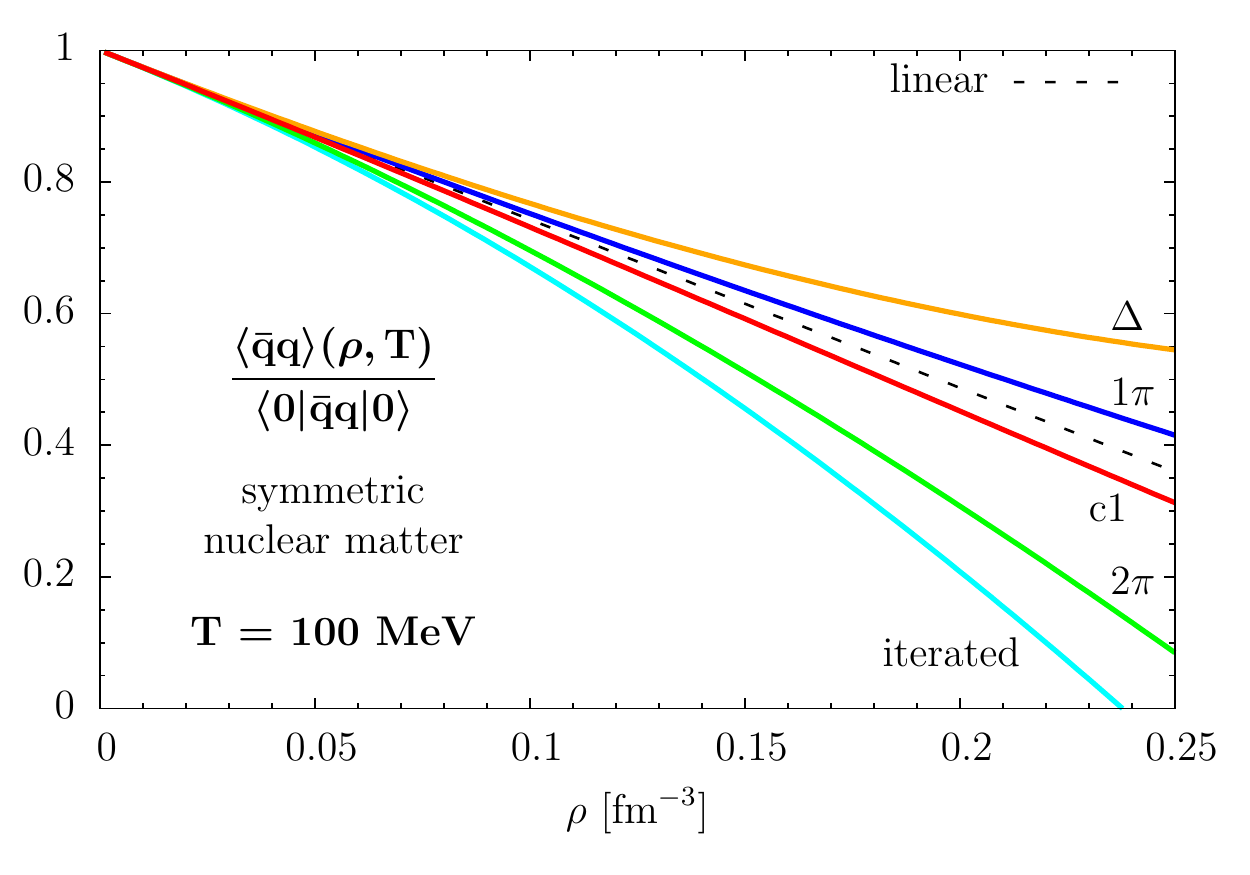}
\caption{Density dependence of the chiral condensate in isospin-symmetric nuclear 
matter taken at a temperature $T = 100$ MeV. Starting from the linear density 
dependence (dashed curve) characteristic of the free nucleon Fermi gas, the 
following interaction contributions are successively added: one-pion exchange Fock 
terms ($1\pi$), second order (iterated) pion exchange, irreducible two-pion 
exchange ($2\pi$), two- and three-body contributions from $2\pi$ exchange with 
intermediate $\Delta(1232)$-excitations ($\Delta$), and two-pion exchange with 
$\pi\pi NN$ vertex involving the chiral low-energy constant $c_1m_\pi^2$. Pauli 
blocking effects are included throughout. }
\label{figure2}
\end{figure}

Fig.\,\ref{figure2} shows a representative example, at  $T = 100$ MeV, displaying 
stepwise the effects of interaction contributions to the density dependence of 
$\langle\bar{q}q\rangle(\rho,T)$ from the chiral two- and three-body kernels 
${\cal K}_{2,3}$. As in the $T=0$ case studied previously \cite{cond}, the pion-mass 
dependence of correlations involving virtual $\Delta(1232)$-excitations turns out 
to be specifically important in delaying the tendency towards chiral symmetry 
restoration as the density increases. Once all one- and two-pion exchange processes 
contributing to ${\cal K}_2$ and ${\cal K}_3$ are added up, the chiral condensate 
at $T=100$ MeV recovers the linear density dependence characteristic of a free 
Fermi gas. However, this recovery is the result of a subtle balance between 
attractive and repulsive correlations and their detailed pion-mass dependences. 
Had we taken into account only iterated one-pion and irreducible two-pion 
exchanges, the system would have become instable not far above normal nuclear 
matter density as seen in Fig.\,\ref{figure2}. In fact, this instability would have 
appeared at even much lower densities in the chiral limit $(m_\pi \rightarrow 0)$. 
This emphasizes once more not only the importance of terms involving $\Delta(1232)
$-excitations, but also the significance of explicit chiral symmetry breaking by 
small but non-zero quark masses in QCD and the resulting physical pion mass, in 
governing nuclear scales.   

\begin{figure}
\includegraphics[totalheight=6.3cm]{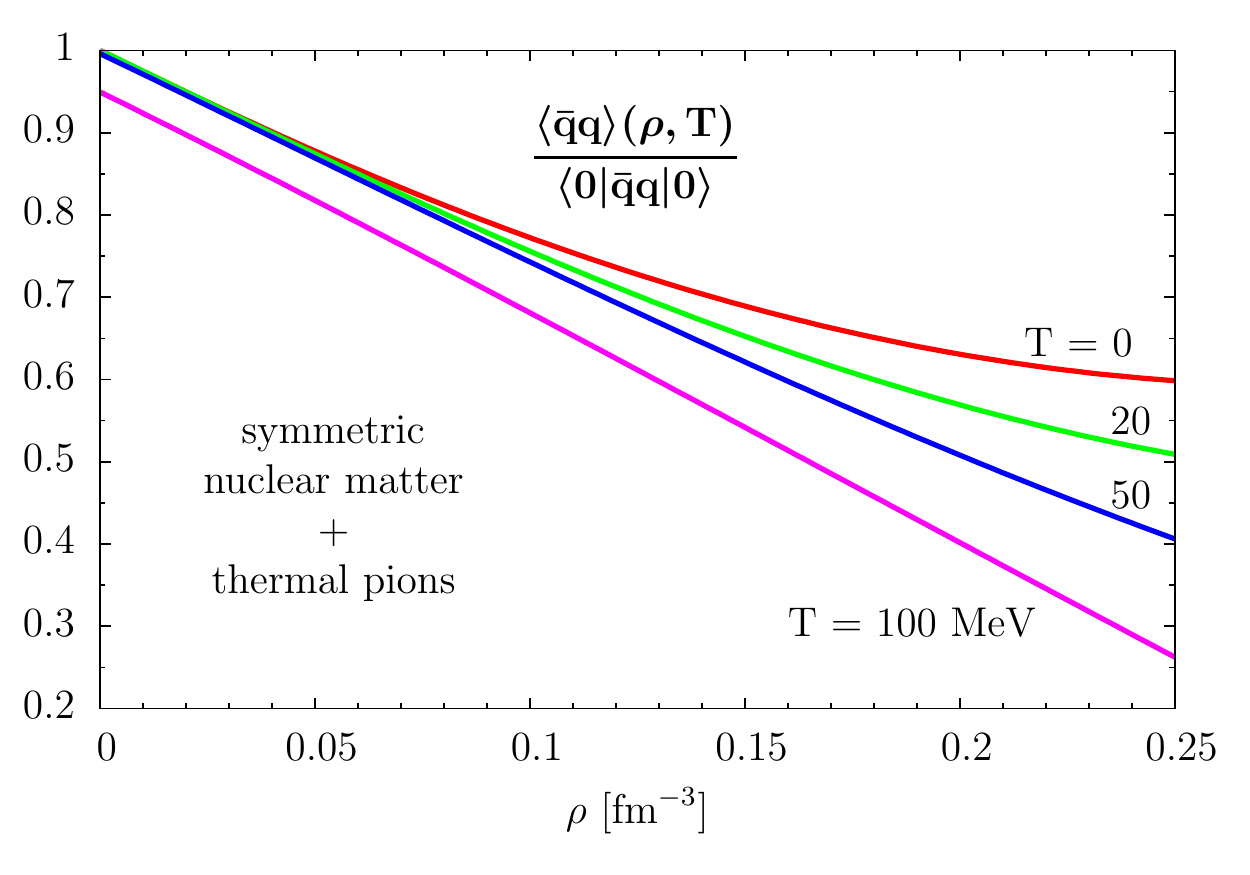}
\caption{Ratio of chiral condensate relative to its vacuum value as function of 
baryon density $\rho$ in symmetric nuclear matter, for different temperatures up 
to $T = 100$ MeV as indicated. The reduction effect due to thermal pions is 
included.}
\label{figure3}
\vspace{-0.5cm}
\end{figure}

Fig.\,\ref{figure3} shows the systematics in the variation of the chiral condensate 
with temperature $T$ and baryon density $\rho$. These results include all nuclear 
correlation effects and also the (small) additional shift arising from thermal 
pions. The latter correction is visible only at the highest temperature considered 
here ($T = 100$ MeV) where the chiral condensate at zero density begins to deviate 
from its vacuum value. According to recent QCD lattice simulations the 
actual crossover transition at which the quark condensate $\langle \bar q q\rangle$ 
drops continuously to zero occurs around a temperature of $T_c \sim 170$ MeV 
\cite{lattice}.  

At zero temperature, the hindrance of the dropping condensate at densities beyond 
normal nuclear matter comes primarily from three-body correlations through 
${\cal K}_3$ which grow rapidly and faster than ${\cal K}_2$ as the density 
increases. The heating of the system reduces the influence of ${\cal K}_3$ 
relative to ${\cal K}_2$ continuously as the temperature rises, so that their 
balance at $T=100$ MeV produces a small net effect in comparison with the free 
Fermi gas.  

In summary, this is the first calculation of the quark condensate at finite 
temperature and density that systematically incorporates chiral two-pion exchange 
interactions in the nuclear medium. Correlations involving intermediate 
$\Delta(1232)$-excitations (i.e. the strong spin-isospin polarizability of the 
nucleon) together with Pauli blocking effects are demonstrated to play a crucial 
role in stabilizing the condensate at densities beyond that of equilibrated nuclear 
matter. The results reported here set important nuclear physics constraints for 
the QCD equation of state at baryon densities and temperatures that are of 
interest e.g. in relativistic heavy-ion collisions. In particular, we find no 
indication of a first order chiral phase transition at temperatures $T\lesssim 
100\,$MeV and baryon densities up to about twice the density of normal nuclear 
matter.   

\end{document}